\documentclass[12pt]{article}
\usepackage{cite}

\begin{document}

\begin{center}
{\bf  Moyal and tomographic probability representations for
f-oscillator quantum states }

\end{center}

\begin{center}
Vladimir I. Man'ko,$^1$ Giuseppe Marmo,$^2$ and Francesco Zaccaria$^2$\\

 $^1$P.N. Lebedev Physical Institute, Leninskii Prospect, 53,
Moscow 119991 Russia\\$^2$Dipartimento di Scienze Fisiche,
Universit\`{a} ``Federico II'' di Napoli and Istituto Nazionale di
Fisica Nucleare, Sezione di Napoli, Complesso Universitario di
Monte Sant Angelo, Via Cintia, I-80126 Napoli, Italy

e-mail:~~manko@sci.lebedev.ru

\end{center}

\begin{abstract}
States of nonlinear quantum oscillators (f-oscillators) are
considered in the Weyl--Wigner--Moyal representation and the
tomographic probability representation, where the states are
described by standard probability distributions instead of wave
functions or density matrices. The evolving integrals of motion
for classical and quantum f-oscillators are found and the solution
for the Liouville equation associated with the probability
distribution on the phase space for this oscillator is obtained
along with the solution of Moyal equation for quantum
f-oscillator, which provide the solutions for partial case of
f-nonlinearity existing in Kerr media. Nonlinear coherent states
and the thermodynamics of nonlinear oscillators are studied.
\end{abstract}

keywords: coherent states, Wigner function, tomographic
probability, deformed oscillator.

\section{Introduction}

The small vibrations of classical and quantum systems are usually
described by linear harmonic oscillator model. Recently
\cite{PhysScr,WignerSym} the notion of f-oscillators and
corresponding nonlinear coherent states  for the oscillator
quantum domain was introduced (see also \cite{Vogel}). The
f-oscillators are nonlinear oscillators with a specific kind of
nonlinearity for which the frequency depends on the oscillation
energy \cite{Salv}. The known q-oscillators \cite{Bid,Mc} are very
particular cases of the nonlinear vibrations with exponential
dependence of the frequency on the energy of the vibrations. The
properties of the f-oscillators and use of these oscillators for
constructing models of different phenomena are intensively
discussed in the literature, see, e.g.,
\cite{Dodonov,UkrainCasimir,Iranian,A,B,C,D,E,F,G,Stern,MendesJPA}.

The nonlinearity of f-oscillators provides a deformation mechanism
creating quantum group structures. For example, the f-deformed
Weyl systems corresponding to the deformation of the
Heisenberg--Weyl group were studied in \cite{Aniello} and
q-nonlinearity corresponds to q-deformed Heisenberg--Weyl group.
Recently in quantum \cite{Mancini96} and classical mechanics
\cite{Olga97,MendesPhysica} the tomographic probability
representation of system states was suggested. In classical
statistical mechanics, the tomographic probability distribution
(tomogram or symplectic tomogram) is the Radon transform
\cite{Radon1917} of the probability distribution of the system on
its phase space which is also the standard probability
distribution. In quantum mechanics, the tomogram is the Radon
transform of the Wigner function \cite{Wig32} of quantum states.
In Moyal approach \cite{Moy49}, the quantum states are described
by the Wigner function and obey the evolution equation which is
similar to the classical Liouville equation. The Wigner function,
which is the Weyl symbol \cite{Weyl} of the density operator
${\rho},$ can take negative values and, in view of this, it is not
a probability distribution. It is called quasidistribution and
contains complete information on quantum states. The symplectic
tomogram also contains complete information on quantum state of a
system but the tomogram is the measurable nonnegative probability
distribution. The approach to describe the quantum states by the
synplectic tomograms is called the tomographic probability
representation of quantum mechanics and its properties are studied
in different aspects in
\cite{Olga,Patrizia,Ventriglia1,Ventriglia2,Ventriglia3,Beppe1,Patrizia1,Patrizia2}.

The aim of this paper is to find integrals of motion and to
consider the behaviour of f-oscillators, both classical and
quantum ones, in the tomographic probability representation. We
study also quantum f-oscillators in the Moyal representation. We
study the thermodynamics of q-oscillator in the case of small
nonlinearity. Both the Weyl--Wigner--Moyal representation and the
tomographic probability representation are examples of realizing
the generic star-product scheme (see, e.g.,
\cite{Olga,Zachos,Fronsdal,Berezin,Ba,Stratonovich}). In this
context, the goal of the paper is also to associate the behaviour
of quantum nonlinear oscillator with the star-product quantization
framework. We point out that the electromagnetic-field vibrations
in Kerr media are a particular example of nonlinear f-oscillations
with nonlinearity function proportional to the field amplitude.

The paper is organized as follows.

In Sect. 2, the properties of f-oscillators are reviewed both in
the classical and quantum domains. In Sect. 3, the Moyal approach
and construction of Weyl symbols, including a deformed Wigner
function, are discussed. In Sect. 4, the tomography of classical
and quantum states based on applying integral Radon transform is
presented. In Sect. 5, nonlinear coherent states are studied. In
Sect. 6, the partition function of nonlinear oscillator is
considered. Finally the prospects and conclusions are done in
Sect. 7.

\section{f-oscillators and kinetic equation}

\subsection{Classical nonlinear oscillator}

The coordinate $x(t)$ of linear harmonic classical oscillator
satisfies the equation of motion
\begin{equation}\label{1}
\ddot{x}+x=0,
\end{equation}
where for simplicity the mass and frequency of the oscillator are
set to be equal to unity, $m=\omega =1$.

For complex amplitude $\alpha (t)= =({1}/{\sqrt{2}})(x+i\dot{x})$
the equation of motion reads
\begin{equation}\label{2}
\dot{\alpha}=-i\alpha,
\end{equation}
and its solutions have the form
\begin{equation}\label{3}
\alpha (t)=\alpha _{0}e^{-it}
\end{equation}
with $\alpha _{0}$ being any initial complex amplitude. The energy
of vibrations is the integral of motion
\begin{equation}\label{4}
E=|\alpha (t)|^{2}=|\alpha
_{0}|^{2}=\frac{\dot{x}^{2}}{2}+\frac{x^{2}}{2}\,.
\end{equation}
Thus the vibration of linear harmonic oscillator has the important
property --- the frequency of the vibration does not depend on the
vibration energy. Another formulation of the same property implies
that the phase of the vibrating linear oscillator is insensitive
to the absolute value of the amplitude of the vibration. The
simplest nonlinearity of vibration just violates this property.
Let us consider the generalization of (\ref{3}) taking the
equation of motion of the form
\begin{equation}\label{EM}
\dot{\alpha}=-i(\alpha \alpha ^{\ast })\alpha.
\end{equation}
One can see that $\frac{d}{dt}(\alpha \alpha ^{\ast })=0$, and the
solutions (\ref{3}) become
\begin{equation}
\alpha (t)=\alpha _{0}e^{-i|\alpha _{0}|^{2}t}.
\end{equation}
Thus for cubic nonlinear equation (\ref{EM}) the frequency of the
vibration depends quadratically on the amplitude of the
vibrations. The generic form of the equation of motion can be
written as
\begin{equation}\label{5}
\dot{\alpha}=-i\omega (\alpha \alpha ^{\ast })\alpha,
\end{equation}
where the frequency function $\omega (E)$ depends on energy
(\ref{4}). The nonlinearity of vibrations is coded by the function
$\omega (E)$. This means that the solution of Eq. (\ref{5})
\begin{equation}\label{6}
\alpha (t)=\alpha _{0}e^{-i\omega (\alpha _{0}\alpha _{0}^{\ast
})t}
\end{equation}
corresponds to the kind of the nonlinearity of the equation. If
one describes the nonlinear-oscillator motion by the Hamiltonian
\begin{equation}\label{7}
H=\alpha _{f}^{\ast }\alpha _{f},
\end{equation}
in which the variable $\alpha _{f}$ reads as
\begin{equation}\label{8}
\alpha _{f}=\alpha f(\alpha ^{\ast }\alpha ),
\end{equation}
the dependence of the frequency on the energy is related to the
dependence on the energy of the real function $f(E)$.

In fact, from (\ref{8}) one has
\begin{equation}\label{9}
\omega (E)=f(E)+Ef^{\prime}(E).
\end{equation}
We call the f-oscillator the classical oscillator with the
Hamiltonian (\ref{7}), where $\alpha _{f}$  is given by (\ref{8}),
thus the oscillator evolves as
\begin{equation}\label{SEM}
\alpha (t)=\alpha _{0}\exp \left\{ -it\left[ f\left( \alpha
_{0}\alpha _{0}^{\ast }\right) +\left( \alpha _{0}\alpha
_{0}^{\ast }\right) f^{\prime}(\alpha _{0}\alpha _{0}^{\ast
})\right] \right\}.
\end{equation}
The Liouville equation for the probability distribution $f(q,p,t)$
on phase-space reads
\begin{equation}\label{LE}
\frac{\partial {\large f}(q,p,t)}{\partial t}+\frac{\partial
{\large f} (q,p,t)}{\partial q}\dot{q}+\frac{\partial {\large
f}(q,p,t)}{\partial p} \dot{p}=0.
\end{equation}
Since for our f-oscillator
\begin{equation}\label{16}
\dot{q}=p\left[ f\left( E\right) +Ef^{\prime }\left( E\right)
\right],\qquad
\dot{p}=-q\left[ f\left( E\right) +Ef^{\prime }\left( E\right) %
\right],
\end{equation}
one has the Liouville equation of the form
\begin{equation}\label{NLE}
\frac{\partial f(q,p,t)}{\partial t}+\frac{\partial f
(q,p,t)}{\partial q}p\left[ f\left( E\right) +Ef^{\prime }\left(
E\right) \right] -\frac{\partial f(q,p,t)}{\partial p}q\left[
f\left( E\right) +Ef^{\prime }\left( E\right) \right] =0,
\end{equation}
where $E=({q^{2}+p^{2}})/{2}=({q^{2}_0+p^{2}_0})/{2}.$

The system of equations (\ref{16}) has two time-dependent
integrals of motion, following the solution (\ref{SEM}) of
equation of motion (\ref{5})
\begin{eqnarray}
&&p_{0}=p\cos \left[ \omega \left( E\right) t\right] +q\sin \left[
\omega \left( E\right) t\right],\label{17}\\
&& q_{0}=-p\sin \left[ \omega \left( E\right) t\right] + q\cos
\left[ \omega \left( E\right) t\right].\label{18}
\end{eqnarray}

For a linear oscillator, $f=1$ and the integrals of motion have
the form \cite{183}
\begin{equation}\label{19}
p_{0}=p\cos t+q\sin t,\qquad q_{0}=-p\sin t+q\cos t
\end{equation}
and their dependence on complex amplitude reads
\begin{equation}\label{20}
\alpha
_{0}=\frac{q_{0}+ip_{0}}{\sqrt{2}}=\frac{q+ip}{\sqrt{2}}e^{it}.
\end{equation}

For f-oscillator, one has the complex integral of motion
generalizing the above invariant
\begin{equation}\label{21}
\alpha _{0}=\frac{q+ip}{\sqrt{2}}\exp \left[ i\omega \left( \frac{%
q^{2}+p^{2}}{2}\right) t\right].
\end{equation}
In terms of the integrals of motion, the solution of Liouville
equation reads
\begin{equation}\label{22}
f\left( q,p,t\right)=f_{0}\Big(q_{0}\left( t\right) ,p_{0}\left(
t\right) \Big),
\end{equation}
where $f_{0}\left( q,p\right) $ is the probability distribution
given at time $t=0$. Thus, given the initial distribution
$f_{0}\left( q,p\right) $ the form of the solution of the kinetic
equation (\ref{NLE}) is
\begin{eqnarray}\label{23}
f\left( q,p,t\right)&=&f_{0}\left\{ \left[ -p\sin \omega \left(
\frac{q^{2}+p^{2}}{2}\right) t+q\cos \omega \left(
\frac{q^{2}+p^{2}}{ 2}\right) t\right] ,\left[ p\cos \omega \left(
\frac{q^{2}+p^{2}}{2}\right) t\right.\right.
\nonumber\\
&&\left.\left.+q\sin \omega \left( \frac{q^{2}+p^{2}}{2}\right)
t\right] \right\}.
\end{eqnarray}

One has the example of classical f-oscillator, the so-called
classical q-oscillator \cite{Bid,Mc}, for which
\begin{equation}\label{24}
f\left( \alpha \alpha ^{\ast }\right) =\sqrt{\frac{\sinh \lambda
\alpha \alpha ^{\ast }}{\lambda \alpha \alpha ^{\ast }}}\ ,\
q=e^{\lambda }.
\end{equation}
For small nonlinearity $\lambda \ll 1$, one has the approximate
value of the nonlinearity function
\begin{equation}\label{25}
f\left( \alpha \alpha ^{\ast }\right) \simeq 1+\frac{\lambda
^{2}}{12}\left( \alpha \alpha ^{\ast }\right) ^{2}=1+\frac{\lambda
^{2}}{12}E^{2},\quad E=\alpha \alpha ^{\ast }
\end{equation}
and, in this case, the frequency of the q-oscillator is
\begin{equation}\label{26}
\omega \left( E\right) =1+\frac{\lambda ^{2}}{4}E^{2},
\end{equation}
so it depends quadratically, for small nonlinearities, on the
energy of vibrations.

\subsection{Quantum nonlinear oscillator}

Let us consider the quantum oscillator annihilation and creation
operators $\hat{a}$  and $\hat{a}^{\dagger }$, respectively. For
Hamiltonian of the form
\begin{equation}\label{27}
\hat{H}=\hat{a}^{\dagger }\hat{a}+\frac{1}{2},\left[
\hat{a},\hat{a}^{\dagger }\right] =\hat{1},
\end{equation}
one has the solution to Heisenberg evolution equation
$\left(\hbar=1\right)$
\begin{equation}\label{28}
\hat{a}(t)=\hat{a}e^{-it},\qquad \hat{a}=\frac{1}{\sqrt{2}} \left(
\hat{q}+i\hat{p}\right) ,
\end{equation}
where $\hat q$ and $\hat p$ are the position and momentum
operators, respectively.

Let us now consider the quantum f-oscillator with the Hamiltonian
\begin{equation}\label{29}
\hat{H}\left( \hat{n}\right) =\frac{1}{2}\left(
\hat{A}_{f}^{\dagger }\hat{A}
_{f}^{{}}+\hat{A}_{f}^{{}}\hat{A}_{f}^{\dagger }\right),\qquad
\hat{n}= \hat{a}^{\dagger }\hat{a},
\end{equation}
where
\begin{equation}\label{30}
\hat{A}_{f}=\hat{a}f\left( \hat{a}^{\dagger }\hat{a}\right) .
\end{equation}

Now we consider the Hamiltonian of the form
\begin{equation}\label{31}
\hat{H}=\hat A_f^\dagger\hat A_f.
\end{equation}
Physical phenomenon where the quantum f-oscillator Hamiltonian
might prove to be useful is the Kerr effect (see, for instance,
\cite{ManWolf}). The description of such a system has been
recently considered in \cite{3530} with the Hamiltonian (in
dimensionless units)
\begin{equation}\label{32}
\hat{H}_{K}=\chi \left( \hat{a}^{\dagger }\right)^{2}
\hat{a}^{2}+\hat{a}^{\dagger }\hat{a},
\end{equation}
where the light frequency is equal to unity and the crystal
nonlinear susceptibility is $\chi$, with the aim of describing the
evolution in terms of the Moyal representation. The above operator
can be easily written as a function of the number operator
$\hat{n}=\hat{a}^{\dagger }\hat{a}$,
\begin{equation}\label{33}
\hat{H}_{K}=\hat{n}+\chi \hat{n}\left( \hat{n}-1\right) .
\end{equation}
This suggests a description by f-oscillators introducing
\begin{equation}\label{34}
\hat{A}_{K}=\hat{a}f_{K}\left( \hat{n}\right)
\end{equation}
with
\begin{equation}\label{35}
f_{K}\left( \hat{n}\right) = \sqrt{1-\chi+\chi\hat{n}}.
\end{equation}
The function $f_{K}$ above introduced allows one to consider
$f_{K}$-nonlinear coherent states and their evolution.

The f-oscillator for both Hamiltonians (\ref{29}) and (\ref{31})
has two time-dependent integrals of motion of the following form:
\begin{equation}\label{36}
\hat{Q}(t)=e^{-i\hat{H}\left( \hat{n}\right)
t}\hat{a}e^{i\hat{H}\left( \hat{n}\right) t},\qquad
\hat{Q}^{\dagger }(t)=e^{-i\hat{H}\left( \hat{n}\right) t}
\hat{a}^{\dagger }e^{i\hat{H}\left( \hat{n}\right) t}.
\end{equation}
The total time derivative of these
operators equals zero. For time $t=0$, the integrals of motion
coincide with the annihilation and creation operators,
respectively.

Due to the structure of the f-oscillator Hamiltonian, the
eigenstates $|n,f>$ of this oscillator Hamiltonian coincide with
the eigenstates of the operator $\hat a^\dagger\hat a|n>$ related
to the usual oscillator. Since the Hamiltonian is expressed in
terms of number operator $ \hat{n}$, the result of commutation
provides an explicit form of the integrals of motion. One can use
the following formulae:
\begin{equation}\label{37}
\hat{a}\varphi \left( \hat{n}\right) =\varphi \left(
\hat{n}+\hat{1} \right) \hat{a}, \qquad \hat{a}^{\dagger }\varphi
\left(  \hat{n}+\hat{1}\right) =\varphi \left( \hat{n}\right)
\hat{a}^{\dagger },
\end{equation}
where $\varphi \left( \hat{n}\right) $ is an arbitrary function.
Then the Hamiltonian (\ref{29}) takes the form $$\hat H\left(
\hat{n}\right) =\frac12\Big(\hat{n}f\left( \hat{ n}\right)
^{2}+\left( \hat{n}+1\right) f\left( \hat{n}+1\right)^{2}\Big).$$
The above integrals of the motion can be written as
\begin{equation}\label{IM}
 \hat{Q}(t)=\hat{a}F\left(
\hat{n},t\right), \qquad \hat{Q}^{\dagger }(t)=F\left(
\hat{n},t\right) \hat{a}^{\dagger },
\end{equation}
where the function $\ F\left( \hat{n},t\right) $ for both
Hamiltonians (\ref{29}) and (\ref{31}) reads
\begin{equation}\label{38}
F\left( \hat{n},t\right) =f\left( \hat{n}\right)\exp\left[
i\left\{ \hat{H} \left( \hat{n}\right) -\hat{H}\left(
\hat{n}-\hat{1}\right) \right\} t\right] .
\end{equation}
The evolution of any initial density state $\rho _{0}\left(
\hat{a}, \hat{a}^{\dagger }\right) $  is given as
\begin{equation}\label{39}
\rho _{{}}\left( \hat{a},\hat{a}^{\dagger },t\right) =\rho
_{0}\left( \hat{Q}(t),\hat{Q}_{{}}^{\dagger }(t)\right).
\end{equation}
The above density operator satisfies the von Neumann evolution
equation
\begin{equation}\label{40}
\frac{\partial }{\partial t}{\rho}+i\left[ \hat{H}\left(
\hat{n}\right) , {\rho}\right] =0.
\end{equation}

\section{Wigner functions}

The Wigner function of the evolving f-oscillator state reads
\begin{equation}\label{41}
W\left( q,p,t\right) =2\mbox{Tr}\left[ \hat{P}\rho \left( t\right)
\exp \left\{ 2\left( \alpha \hat{a}^{\dagger }-\alpha ^{\ast
}\hat{a}\right) \right\}\right],
\end{equation}
where $\hat{P}$  is the parity operator $\hat{P}=\left( -1\right)
^{\hat{a}^{\dagger }\hat{a}}$ and complex number $\alpha $ is
expressed in terms of the position and momentum as $\alpha
=({q+ip})/{\sqrt{2}}$. The parity operator commutes with the
Hamiltonian. In view of this, the Wigner function takes the form
\begin{equation}\label{W}
W\left( q,p,t\right) =2\mbox{Tr}\left[ \hat{P}\rho _{0}\exp\left\{
2\left( \alpha \hat{Q}^\dagger(t)-\alpha ^{\ast }\hat{Q}(t)\right)
 \right\}\right].
\end{equation}
The Wigner function (\ref{W}) is the solution to the Moyal
evolution equation \cite{Moy49} for the f-oscillator. To calculate
explicitly the Wigner function, one needs to evaluate the trace in
Eq. (\ref{38}). It is worth noting that in Eq. (\ref{W}) the Weyl
system operator has the form of f-deformed shift operator which
was used in \cite{Aniello}. The deformation function is given by
Eq. (\ref{38}).

When the initial state is the ground oscillator state $\rho
_{0}=|0><0|$, one has the Wigner function
\begin{equation}\label{42} W_{0}\left( q,p,t\right) =2<0|\exp \left[
2\left( \alpha \hat{Q}^{\dagger }(t)-\alpha ^{\ast
}\hat{Q}(t)\right) \right] |0>.
\end{equation}
Thus the Wigner function is proportional to the expectation value
of the deformed shift operator in the vacuum state.

Let us now discuss the deformed creation and annihilation
operators $\hat{A}_{f}^{\dagger }$ and $\hat{A}_{f}$ and the
corresponding deformed parity operator, more formally.

The replacement of the usual boson operators $\hat{a}^{\dagger }$
and $\hat{a}$ by the deformed ones gives the deformation of the
shift operator
\begin{equation}\label{43}
\exp \left[ \left( \alpha \hat{a}^{\dagger }-\alpha ^{\ast
}\hat{a}\right) \right] \rightarrow \exp \left[ \left( \alpha
\hat{A} _{f}^{\dagger }-\alpha ^{\ast }\ \hat{A}_{f}^{{}}\right)
\right],
\end{equation}
and the deformation of parity operator
\begin{equation}\label{44}
\left( -1\right) ^{\hat{a}^{\dagger }\hat{a}}=\exp \left( i\pi
\hat{a} ^{\dagger }\hat{a}\right) \rightarrow \exp \left( i\pi
\hat{A} _{f}^{\dagger }\hat{A}_{f}\right) .
\end{equation}
Then one has the possibility of introducing deformed Wigner
functions using f-deformation of star-product quantization scheme.
Namely, one gets either
\begin{equation}\label{45}
W_{f}^{(1)}(q,p)=2\mbox{Tr}\left[\hat{P}\hat\rho\exp\left\{
2\left( \alpha \hat{A} _{f}^{\dagger }{}^{{}}-\alpha ^{\ast }
\hat{A}_{f}^{{}}\right) \right\} \right],
\end{equation}
keeping the parity operator in the usual form or
\begin{equation}\label{46}
W_{f}^{(2)}(q,p)=2\mbox{Tr}\left[ \exp \left( i\pi
\hat{A}_{f}^{\dagger }\hat{A} _{f}\right)\hat\rho \exp
\left\{2\left( \alpha \ \hat{A}_{f}^{\dagger }-\alpha ^{\ast }
\hat{A}_{f}^{{}}\right) \right\} \right] .
\end{equation}
For small deformations, the Wigner functions we have introduced do
not differ much from the standard Wigner quasidistributions. The
evolution in time of the deformed Wigner functions can be obtained
by using the integrals of motion (\ref{IM}).

\section{Symplectic tomography}

Let us discuss in this section symplectic tomography
\cite{Mancini96} of classical states and nonlinear quantum states.

We will focus on symplectic tomograms of f-oscillator states.

Given a density state ${\rho}$, the symplectic tomogram also
called tomographic symbol of the density operator is defined as
\begin{equation}\label{47}
w\left( X,\mu ,\nu \right) =\mbox{Tr}\left[{\rho}\delta \left(
X\hat{1 }-\mu \hat{q}-\nu \hat{p}\right)\right],
\end{equation}
where $X,\mu ,\nu $  are real numbers while $\hat{1},$ $\hat{q} $
and $\hat{p}$ are identity, position and momentum operators,
respectively. In terms of the Wigner function, the tomogram reads
\begin{equation}\label{48}
w\left( X,\mu ,\nu \right) =\int W\left( q,p\right) \delta \left(
X-\mu q-\nu p\right) \frac{dqdp}{2\pi }.
\end{equation}
The tomogram is a probability distribution function of the
position $X$, i.e.,
\begin{equation}\label{49}
w\left( X,\mu ,\nu \right) \geq 0
\end{equation}
and
\begin{equation}\label{50}
\int w\left( X,\mu ,\nu \right) dX=1.
\end{equation}
The parameters $\mu =s\cos \theta $ and  $\nu =s^{-1}\sin \theta $
correspond to a reference frame in the phase space which is scaled
$\left( q\rightarrow sq, p\rightarrow s^{-1}p\right) $ and
thereafter rotated $\left( sq\rightarrow qs\cos \theta
+ps^{-1}\sin \theta \right).$

In classical mechanics, an analog of the tomographic symbol can be
constructed for an arbitrary observable $\mathcal{A}\left(
q,p\right)$,  which is a function on the phase space, as the Radon
integral
\begin{equation}\label{51}
w_{\mathcal{A}}\left( X,\mu ,\nu \right) =\int \mathcal{A} \left(
q,p\right) \delta \left( X-\mu q-\nu p\right) dq\,dp.
\end{equation}
Let us discuss now the tomographic representation for the
classical f-oscillator. The solution to the Liouville equation
(\ref{NLE}) has the following tomogram for the
nonlinear-oscillator state:
\begin{eqnarray}\label{52}
&&w\left( X,\mu ,\nu ,t\right) =\int f_{0}\left\{ \left[ -p\sin
\omega \left( \frac{q^{2}+p^{2}}{2}\right) t\right.\right.\nonumber\\
&&\left.\left. +q\cos \omega \left( \frac{q^{2}+p^{2}}{2}\right)
t\right] ,\left[ p\cos \omega \left(
\frac{q^{2}+p^{2}}{2}\right) t+q\sin \omega \left( \frac{q^{2}+p^{2}}{2}%
\right) t\right]\right\}\nonumber\\
&&\times\delta \left( X-\mu q-\nu p\right) dq\,dp.
\end{eqnarray}
The integral provides the solution (the tomogram) in terms of
initial value of the tomogram $w_0\left( X,\mu ,\nu \right) $
which is given as
\begin{equation}\label{53}
w_{0}\left( X,\mu ,\nu \right) =\int f_{0}\left( q,p\right) \delta
\left( X-\mu q-\nu p\right) dq\,dp.
\end{equation}
The symplectic tomogram of the f-oscillator quantum states is
given in terms of the density operator (\ref{39}) as
\begin{equation}\label{54}
w\left( X,\mu ,\nu,t \right) =\mbox{Tr}\left[\rho_0\left(\hat
Q(t), \hat Q^\dagger (t)\right)\delta \left( X-\mu\hat q-\nu\hat
p\right)\right].
\end{equation}

The quantum tomogram provides the probability distribution of the
f-oscillator position $w(x,1,0)=P(x)$ and the f-oscillator
momentum $w(p,0,1)={\cal P}(p)$.

The ground state of the nonlinear oscillator satisfies the
equation $A_f|0,f>=0$, which also satisfies the equation for the
usual oscillator ground state $a|0,f>=0$. This means that the
tomogram of the f-oscillator ground state reads
\begin{equation}\label{55}
w_0(X,\mu ,\nu) =\mbox{Tr}\left[|0,f><0,f|\delta\left(X-\mu\hat
q-\nu\hat
p\right)\right]=\frac{1}{\sqrt{\pi\left(\mu^2+\nu^2\right)}}
\exp\left(-\frac{X^2}{\mu^2+\nu^2}\right).
\end{equation}
The excited states of the nonlinear f-oscillator $|n,f>$ have the
tomogram, which also coincides with the tomogram of the excited
state $|n>$ of the usual harmonic oscillator, which reads
\begin{equation}\label{56}
w_n(X,\mu ,\nu) =w_0(X,\mu,\nu)\frac{1}{2^nn!}
H_n^2\left(\frac{X^2}{\sqrt{\mu^2+\nu^2}}\right).
\end{equation}

The constructions of Wigner functions and tomograms of of the
nonlinear oscillator's quantum states uses the operators
(dequantizers) which are ingredients of star-product quantization
schemes. Thus we found the symplectic tomograms of the basis
excited states for both harmonic and nonlinear oscillators which
turn out to be the same probability distributions.

\section{f-oscillator coherent states}

In \cite{PhysScr,WignerSym} the notion of f-oscillator nonlinear
coherent state was introduced (see also \cite{Vogel}). The states
are constructed as eigenstates of the operator
$\hat{A}_f=\hat{a}f\left( \hat{n} \right) $,  with
$\hat{n}=\hat{a}^{\dagger }\hat{a}$ and the equation
\begin{equation}\label{57}
\hat{A}_f|\alpha ,f>=\alpha |\alpha ,f>,
\end{equation}
where $\alpha $ is a complex number. For $f\left( \hat{n}\right)
=1$, the nonlinear coherent state coincides with usual coherent
state $|\alpha >$, i.e.,
\begin{equation}\label{58}
|\alpha >=\exp \left[ -\frac{|\alpha |^{2}}{2}\right]
\sum_{n=0}^{\infty }\frac{\alpha ^{n}}{\sqrt{n!}}|n>
\end{equation}
 and
$\hat{a}^{\dagger }\hat{a}|n>=n|n>,\quad n=0,1,2,\ldots$

The nonlinear coherent state is expressed as the following series
in the Fock number states $|n>$
\begin{equation}\label{59}
|\alpha ,f>=N_{f}\sum_{n=0}^{\infty }\frac{\alpha^{n}}{f\left(
n\right)!\sqrt{n!}}|n>,\quad f\left( n\right) !=f\left( 0\right)
f\left( 1\right) f\left( 2\right) \ldots f\left( n-1\right)
f\left( n\right) ,
\end{equation}
where the normalization is given by the series
\begin{equation}\label{60}
N_{f}=\left( \sum_{n=0}^{\infty }\frac{|\alpha
|^{2n}}{\left[f\left( n\right)!\right]^2n!} \right) ^{-{1}/{2}}.
\end{equation}

In the position representation, the nonlinear coherent state reads
\begin{equation}
\psi _{\alpha ,f}\left( x\right) =N_{f}\sum_{n=0}^{\infty
}\frac{\alpha ^{n} }{f\left( n\right) !\sqrt{n!}}\left[
e^{-\frac{x^{2}}{2}}\frac{1}{\pi^{1/4}\sqrt{n!2^{n}}}H_{n}\left(
x\right) \right] .
\end{equation}

One can introduce two-mode nonlinear coherent states of the
two-dimensional  f-oscillator, for example, from operators
\begin{equation}\label{61}
\hat{A}_{1}=\hat{a}_{1}f\left( \hat{n}_{1},\hat{n}_{2}\right)
\quad\mbox{ and}\quad\hat{A}_{2}=\hat{a}_{2}f\left(
\hat{n}_{1},\hat{n}_{2}\right) ,
\end{equation}
where $\hat{a}_{1}$ and $\hat{a}_{2}$ are annihilation operators
for linear oscillators and $\hat{n}_{i}=\hat{a}_{i}^{\dagger
}\hat{a} _{i}$, $i=1,2.$ Their eigenstates are defined by the
equations
\begin{equation}\label{62}
\hat{A}_{i}|\alpha _{1}\alpha _{2},f>=\alpha _{i}|\alpha
_{1}\alpha _{2},f>.
\end{equation}
The nonlinearity is coded by the function of two variables
$f\left( \hat{ n}_{1},\hat{n}_{2}\right) $ which, in fact, depends
on the energies of both mode vibrations.

As an example, let us consider the nonlinearity given by the one
variable function $f\left( \hat{n}_{1}+\hat{n}_{2}\right) $ where
the coherent states are entangled by construction. The explicit
form of such nonlinear oscillator coherent state in the position
representation reads
\begin{equation}\label{63}
\Psi _{\alpha _{1}\alpha _{2}f}\left( x,y\right)
=N_{f}\sum_{n_{1}n_{2}=0}^{\infty }\frac{\alpha _{1}^{n_{1}}\alpha
_{2}^{n_{2}}}{\sqrt{n_{1}!n_{2}!}f\left( n_{1}+n_{2}\right) !}
<xy|n_{1}n_{2}>.
\end{equation}
The normalization constant is given by the expression
\begin{equation}\label{64}
N_{f}=\left( \sum_{n_{1}n_{2}=0}^{\infty }\frac{|\alpha
_{1}|_{{}}^{2n_{1}}|\alpha
_{2}|_{{}}^{2n_{2}}}{n_{1}!n_{2}!\left[f\left( n_{1}+n_{2}\right)
!\right]^2}\right) ^{-{1}/{2}}
\end{equation}
and
\begin{equation}\label{65}
<xy|n_{1}n_{2}>=\frac{e^{-\frac{x^{2}}{2}-\frac{y^{2}}{2}}}{\sqrt{\pi
}\sqrt{ 2^{n_{1}+n_{2}}n_{1}!n_{2}|}}H_{n_{1}}\left( x\right)
H_{n_2}\left( y\right) .
\end{equation}
The functions $H_{n_{i}}$ are Hermite polynomials. For $f=1$, the
nonlinear coherent state becomes separable two-mode coherent state
\begin{equation}\label{66}
\Psi _{\alpha _{1}\alpha _{2}}\left( x,y\right) =\frac{e^{-\frac{x^{2}}{2}-%
\frac{y^{2}}{2}}}{\sqrt{\pi }}\exp \left[ -\frac{|\alpha _{1}|^{2}}{2}-\frac{%
|\alpha _{2}|^{2}}{2}+\sqrt{2}\left( \alpha _{1}x+\alpha _{2}y\right) -\frac{%
\alpha _{1}{}^{2}}{2}-\frac{\alpha _{2}{}^{2}}{2}\right] .
\end{equation}

Let us discuss the introduced entanglement in more details.

The pure state of two-mode system is separable if its wave
function is factorized, namely,
\begin{equation}\label{67} \Psi \left(
x_{1},x_{2}\right) =\psi _{1}\left( x_{1}\right)\psi _{2}(x_{2}),
\end{equation}
where we denote the modes coordinates as $x_{1}$ and $x_{2}$. A
wave function, which is the superposition of separable states, is
entangled.

The usual nondeformed two-mode coherent states are separable. In
fact,
\begin{equation}\label{68}
|\alpha _{1}\alpha _{2}>=|\alpha _{1}>|\alpha _{2}> \rightarrow
<x_{1}x_{2}|\alpha _{1}\alpha _{2}>=<x_{1}|\alpha
_{1}><x_{2}|\alpha _{2}>,
\end{equation}
where
\begin{equation}\label{69}
<x_{1}|\alpha _{1}>=\exp \left[ -\frac{\left| \alpha _{1}\right| ^{2}}{2}%
\right] \sum_{n_{1}=0}^{\infty }\frac{\alpha _{1}^{n_{1}}e^{-\frac{%
x_{1}^{2}}{2}}}{\pi^{1/4}\sqrt{n_{1}!}\sqrt{2^{n_{1}}n_{1}!}}%
H_{n_{1}}\left( x_{1}\right)
\end{equation}
 and
\begin{equation}\label{70}
<x_{2}|\alpha _{2}>=\exp \left[ -\frac{\left| \alpha _{2}\right| ^{2}}{2}%
\right] \sum_{n_{2}=0}^{\infty }\frac{\alpha _{2}^{n_{2}}e^{-\frac{%
x_{2}^{2}}{2}}}{\pi^{1/4}\sqrt{n_{2}!}\sqrt{2^{n_{2}}n_{2}!}}%
H_{n_{2}}\left( x_{2}\right) .
\end{equation}

The nonlinear coherent states which we have introduced have the
structure
\begin{equation}\label{71} |\alpha _{1}\alpha
_{2},f>=N_{f}\sum_{n_{1}n_{2}=0}^{\infty }\frac{\alpha
_{1}^{n_{1}}\alpha _{2}^{n_{2}}}{\sqrt{n_{1}!n_{2}!}f\left(
n_{1}+n_{2}\right) !}|n_{1}n_{2}>.
\end{equation}

In the case of $f\left( n_{1}+n_{2}\right) =1$,  this series is
reduced to the product of two series, each of them depending on
single variable. But this is not the case in presence of the
nonlinear function $f$  and therefore the  nonlinearity creates
entanglement.

\section{Thermodynamics of f-oscillators}

For standard harmonic oscillators with Hamiltonian (\ref{27}), the
thermodynamical properties are determined by the partition
function
\begin{equation}\label{72}
Z_{0}\left( \beta \right) =\mbox{Tr}\,
e^{-\beta\hat{H}}=\sum_{n=0}^{\infty }e^{-\beta \left(
n+{1}/{2}\right) }=\frac{1}{2\sinh({\beta }/2)}\,.
\end{equation}
The other characteristics are given using the partition function,
like the energy
\begin{eqnarray}\label{73}
&&E=\mbox{Tr}\left(\hat{H}\frac{e^{-\beta \hat{H}}}{Z_{0}\left( \beta \right) }\right)=%
\frac{1}{Z_{0}\left( \beta \right) }\sum_{n=0}^{\infty }\left( n+\frac{1}{2}%
\right) e^{-\beta \left( n+{1}/{2}\right) }\nonumber\\
&&=\frac{1}{2}+\frac{1}{%
Z_{0}\left( \beta \right) }\left( -\frac{\partial }{\partial \beta }%
Z_{0}\left( \beta \right) \right)
\end{eqnarray}
and the entropy
\begin{eqnarray}\label{74}
&&S=-\mbox{Tr}\left( \frac{e^{-\beta \hat{H}}}{Z_{0}\left( \beta \right) }\log \frac{%
e^{-\beta \hat{H}}}{Z_{0}\left( \beta \right) }\right)\nonumber\\
&& =-\frac{1}{Z_{0}\left( \beta \right) }\sum_{n=0}^{\infty
}e^{-\beta \left( n+{1}/{2}\right)}\left[ -\beta \left(
n+\frac{1}{2}\right)
-\log Z_{0}\left( \beta \right) \right]\nonumber\\
&& =\beta E+\log Z_{0}\left( \beta \right),
\end{eqnarray}
which gives
\begin{equation}\label{75}
E-TS=-T\log Z_{0}\left( \beta \right),
\end{equation}
which is the free energy.

To consider the thermodynamics for f-oscillators, let us use the
Hamiltonian $\hat{H}_{f}=\hat{A}_{f}^{\dagger
}\hat{A}_{f}+{1}/{2}$. The partition function for such a system
will be denoted $Z_{f}\left( \beta \right) $. For the nonlinearity
function containing a small deviation from the linear case, we
assume the nonlinearity to be expressed as
\begin{equation}\label{76}
f\left( \hat{n}\right) \approx \left( 1+g\psi \left(
\hat{n}\right) \right),
\end{equation}
where $g\ll 1$  and $\psi \left( \hat{n}\right) $  is some
function of number of vibrations, i.e., we assume
\begin{equation}\label{77}
\hat{A}_{f}\approx \hat{a}\left[ 1+g\psi \left( \hat{n}\right)
\right].
\end{equation}
Then the proposed Hamiltonian of nonlinear oscillator takes the
form
\begin{equation}\label{78}
\hat{H}_{f}=\hat{A}_{f}^{\dagger }\hat{A}_{f}+\frac{1}{2}\approx \hat{a}%
^{\dagger }\hat{a}+\frac{1}{2}+g\chi \left( \hat{n}\right),
\end{equation}
in which
\begin{equation}\label{79}
\chi \left( \hat{n}\right) =2\psi \left( \hat{n}\right).
\end{equation}

Now one can calculate the partition function for this f-oscillator
\begin{equation}\label{80}
Z_{f}\left( \beta \right) =\mbox{Tr}\, \exp \left[ -\beta \left( \hat{a}%
^{\dagger }\hat{a}+\frac{1}{2}\right) -\beta g\chi \left( \hat{n}\right) %
\right].
\end{equation}
Keeping the same accuracy, i.e., linear terms in the small
parameter $g$, we get
\begin{equation}\label{81}
Z_{f}\left( \beta \right) =\sum_{n=0}^{\infty }\exp\left[ -\beta
\left( n+\frac{1}{2}\right) \right] \left[ 1- \beta g\chi \left( \hat{%
n}\right)\right].
\end{equation}
Repeating the calculations as done for the linear oscillator, we
get the deformed partition function
\begin{equation}\label{82}
Z\left( \beta \right) _{f}\approx Z_{0}\left( \beta \right) \left[
1-\beta g<\chi \left( \hat{n}\right) >\right],
\end{equation}
where
\begin{equation}\label{83}
<\chi \left( \hat{n}\right) >=\frac{1}{Z_{0}\left( \beta \right) }%
\sum_{n=0}^{\infty }\chi \left( n\right) e^{-\beta \left(
n+{1}/{2}\right) }.
\end{equation}
Having the correction to the linear oscillator partition function,
one can compute the small corrections to all the thermodynamical
characteristics of the nonlinear f-oscillator, like free energy,
entropy, etc.

For example, for quantum q-oscillator considered in \cite{Aniello}
\begin{equation}\label{84}
f\left( \hat{n}\right) =\sqrt{\frac{\sinh \lambda \hat{n}}{\lambda \hat{n}}}%
\approx 1+\frac{\lambda ^{2}}{12}\hat{n}^{2}\,,
\end{equation}
 we have
\begin{equation}\label{85}
\chi \left( \hat{n}\right) =\frac{\lambda ^{2}}{6}\hat{n}^{2}.
\end{equation}
Here $\lambda $ is a small parameter related to $g$,  i.e.,
$g={\lambda ^{2}}/{6}$. So in the present case, we have
\begin{equation}\label{86}
<\chi \left( \hat{n}\right) >=<\hat{n}^{2}> = \frac{1}{Z_{0}
\left(\beta \right) }\sum_{n=0}^{\infty} n^{2}e^{-\beta \left(
n+{1}/{2}\right) }=\frac{1}{2}\,\coth\frac{\beta}{2} \left(
\coth\frac{\beta }{2}-1\right).
\end{equation}
Thus the correction for the partition functions for small
q-oscillator nonlinearity reads
\begin{equation}\label{87}
Z\left( \beta \right)_{f}-Z_{0}\left( \beta \right) \approx
-\frac{g\beta}{2}\,\coth\frac{\beta}{2}\left(
\coth\frac{\beta}{2}-1\right).
\end{equation}
For large temperature $\beta\to 0$, the thermodynamic
characteristics of the nonlinear quantum oscillator become the
thermodynamic characteristics of the classical q-oscillator.

\section{Conclusions}

To conclude, we point out the main results of our work.

We studied nonlinear f-oscillators in both classical and quantum
settings and found new time-dependent integrals of motion for the
oscillators. For both settings, we obtained the solutions to the
kinetic equations like Liouville classical equation for
probability distribution on the phase space and von Neumann
equation for the density operator of the f-oscillator states.
Using the deformation function determining the f-oscillator, we
introduced the deformed Wigner functions by deforming Weyl
displacement operator in the formalism of Moyal star-product
quantization. Also we studied symplectic tomography of the
f-oscillator states and obtained the tomographic probability of
the ground and excited states of the f-oscillators, which were
shown to coincide with the tomograms of the excited states of the
linear harmonic oscillator. We showed that the particular
nonlinearity of vibrations we have considered in the case of
two-mode nonlinear coherent states has created entangled states.
We calculated the thermodynamic properties of the nonlinear
quantum oscillators in the case of small nonlinearity and found
explicit corrections to the partition function of the linear
oscillator induced by small nonlinearity of q-oscillator. We
showed that known Kerr nonlinearity of media can be considered
within the framework of a specific nonlinear oscillator formalism.

\section*{Acknowledgements}

VIM thanks Dipartimento di Scienze Fisiche, Universit\`{a}
``Federico II'' di Napoli and Istituto Nazionale di Fisica
Nucleare, Sezione di Napoli for kind hospitality and the Russian
Foundation for Basic Research for a partial support under Projects
Nos.~07-02-00598 and 09-02-00142.

\end{document}